\PassOptionsToPackage{square,numbers,sort&compress}{natbib}
\documentclass{IET-Conf-Paper}
\usepackage{tabularx}
\usepackage{makecell}
\usepackage{multirow}
\usepackage{hyperref}

\header{24\textsuperscript{the} Wind \& Solar Integration Workshop | Berlin, Germany | 07-10 October 2025}

\usepackage{subfiles} 
\usepackage{booktabs, tabularx, makecell, array, multirow}
\newcolumntype{Y}{>{\raggedright\arraybackslash}X}

\usepackage{pifont} 

\usepackage{graphicx}
\usepackage{subcaption}
\usepackage{float}

\usepackage{todonotes}

\usepackage{
    fix-cm
} 

\begin{document}
    \title{Exploring the impacts of demand scenarios, weather variability and mitigation
    of emissions on Morocco’s hydrogen market and renewable transition pathways}

    \author{Estefanía Duque Pérez\ad{1}\ad{*}, Lukas Jansen\ad{1}, Benedikt
    Haeckener\ad{1}}

    \address{\add{1}{Fraunhofer IEE, Fraunhofer Institute for Energy Economics and Energy System Technology, Kassel, Germany},
    \add{*}{estefania.duque.perez@iee.fraunhofer.de} }

    \vspace{1em}
    \textbf{Note:} This paper is a preprint of a paper accepted in the proceedings
    of 24th Wind \& Solar Power System Integration Workshop (WISO 2025) and is subject
    to Institution of Engineering and Technology Copyright. When the final
    version is published, the copy of record will be available at the IET Digital
    Library.

    \keywords{Sector-coupled energy system, ammonia, methanol, green steel,
    green hydrogen, new value chains}

    \begin{abstract}
        The global demand for green hydrogen and its derivatives is growing rapidly
        as a cornerstone for decarbonizing hard-to-abate sectors. Morocco, endowed
        with abundant solar and wind resources, ambitions to capture up to 4\% of
        the global PtX market by 2030, positioning itself as a strategic partner
        for Europe’s energy transition. Yet, uncertainty persists regarding
        European demand trajectories, infrastructure readiness, and investment
        risks. This study evaluates Morocco’s hydrogen transition through 2035 using
        a sector-coupled capacity expansion model. We compare industry
        reallocation and hydrogen export-oriented scenarios, assessing their impacts
        under interannual weather variability and financial sensitivities. Both
        scenarios require a tripling of current renewable and electrolyzer
        capacities, with hydrogen demand reaching approximately up to 38 TWh by 2035.
        Lower financing costs (WACC) have a greater effect on system costs and competitiveness
        than stricter CO$_{2}$ constraints or weather variability. The trade-off
        between domestic energy security and export competitiveness is pronounced,
        but both pathways are technically feasible and aligned with. These findings
        provide evidence-based guidance for policymakers to balance Morocco’s domestic
        and export ambitions in the evolving hydrogen market.
    \end{abstract}

    \maketitle

    \section{Introduction}
    
    Power-to-X (PtX) fuels, green hydrogen and its derivatives such as ammonia
    and e-methanol, are widely viewed as essential for decarbonizing hard-to-electrify
    industries and long-distance transport, while also providing flexible back-up
    capacity in power systems \cite{IEA_GlobalHydrogenReview_2022}. The direction
    of travel is clear: hydrogen and PtX will underpin the resilience and
    climate neutrality goals of Europe and other countries, such as Japan. Yet
    uncertainties persist regarding demand volumes, choice of carriers, and
   
    delivery routes, which in turn create volatility in global market formation \cite{ECA_RenewableH2_2024}.

    Morocco occupies a distinctive position in this emerging global PtX landscape.
    It's energy system is characterized by a high dependence on fossil fuel
    imports around 90\% of its primary needs from abroad \cite{WorldBank_WDI_EnergyImports_Morocco_2023},
    which exposes the country to external price volatility and supply risks
    \cite{IEA_Morocco_2019}. Yet benefits from abundant wind and solar resources,
    strategic industrial ties to the EU, and access to maritime and grid interconnections
    \cite{pfennig2023global, TradingEconomics_Morocco_GDP_Sector_2024,
    TradeGov_Morocco_CCG_2024}.

    Building on these advantages, Morocco has moved early with an increasingly
    proactive national strategy: the 2021 Green Hydrogen Roadmap and the 2024 “Morocco
    Offer” frame large-scale PtX deployment, including land provision, governance
    reforms, and incentives for export-oriented hydrogen and ammonia projects
    \cite{MoroccoOfferCircular2024EN, CMS_Morocco_H2_Law_2024}. This is reinforced
    by bilateral partnerships with Germany and the EU \cite{BMWK_H2StrategyUpdate_2023}
    and a growing pipeline of PtX projects \cite{BakerMcKenzie_Morocco_Offer_2024}.

    Morocco's PtX development hinges on external demand, especially Europe's off-take
    clarity, and on domestic enablers such as financial de-risking and cost-competitive
    supply routes \cite{FraunhoferISE2023_PtXImports}. This creates a dual
    challenge: shifting from hydrocarbon and ammonia importer to exporter of green
    molecules while ensuring domestic value creation and reducing exposure to external
    demand volatility \cite{TradeGov_Morocco_CCG_2024}.

    Balancing exports with domestic needs remains essential. Morocco must
    decarbonize cement, petrochemicals, and heavy transport, electrify households,
    and manage system stability, water constraints, and social acceptance. Recent
    studies suggest that ambitious renewable pathways are only slightly more costly
    than least-cost options while delivering substantial emissions reductions\cite{Slimani2024};
    aligning hydrogen production with renewable profiles can also lower electricity
    costs and support competitive exports \cite{Schumm2025}.

    Despite the momentum, a knowledge gap persists. Previous studies applying sector-coupled
    models and open-source energy frameworks such as OSeMOSYS \cite{Slimani2024}
    have analyzed Morocco's renewable transition, but they generally focus on
    aggregate system-level outcomes rather than explicit optimization of green
    hydrogen production or its downstream derivatives. Recent PyPSA-Earth based case
    studies on Morocco analysed different export pathways for Morocco and their impact
    on the domestic electricity market \cite{Schumm2025, jalbout2025techno}.
    However, a systematic comparison of hydrogen-based pathways considering different
    domestic use and export scenarios, as well as direct electrification and
    efficiency measures across critical subsectors at high resolution is still
    lacking.


    Given that industry contributes 24-25\% of GDP including phosphates, fertilizers,
    automotive, and aerospace, there is significant potential for both domestic
    hydrogen use and new value chains such as green ammonia and green steel
    \cite{ADBG_Morocco_2024}.

    This study addresses this gap through a capacity expansion and dispatch
    model that integrates green hydrogen supply chains under water and land-use
    constraints. We develop scenarios contrasting hydrogen export-oriented and domestically
    or high-value industry reallocation focused pathways, combined with
    sensitivities on CO$_{2}$ mitigation ambitions and weather year uncertainty.

    The analysis provides evidence-based insights to support coherent policies and guide investment decisions 
    for Morocco's hydrogen and renewable transition, balancing domestic
    decarbonization with international competitiveness and resilience.

    \section{Methodology}
    
  This section describes the integration of three PtX supply chains: green steel,
  methanol, and ammonia for domestic use within a high-resolution energy system
  model analysis. The methodology encompasses demand pathway projections,
  potential area analysis, scenario definitions, and welfare proxy calculations.

  \subsection{Model framework and workflow}

  For this study, we employ and extend the sector-coupled version of the global
  PyPSA-Earth model \cite{parzen2023pypsa, Pypsasec2025}, an open-source, multi-regional,
  hourly-resolved, multi-energy system optimization model workflow that
  integrates multiple energy and non-energy sectors at high spatial resolution. The
  model minimizes total system costs, including both investment and operational
  expenses, subject to technical and policy constraints. The model workflow is
  applied to assess the medium-term implications of two main scenarios tailored
  to Morocco's current context.

  The main objective is to evaluate the impact of integrating extended PtX value
  chains on energy demands, technology capacity expansions, and on a simplified
  welfare calculation under the two proposed scenarios. Through sensitivity analysis,
  we furthermore assess the implications of uncertainties on results to provide
  futher insigts for informed decision-making related to different strategic
  priorities.

  We extend the PyPSA-Earth sector-coupled model with several key enhancements
  to better represent Morocco's emerging role in power-to-X (PtX) value chains
  and the technical constraints associated with renewable hydrogen certification.
  The modular architecture of PyPSA-Earth allows for extensive configurability through
  user-defined data channels, facilitating adaptation to the system under study.
  Furthermore, the open-source PyPSA community has eased the implementation of
  the extensions with comparable modelling expercises \cite{Neumann2025}.

  Key features added for this analysis to improve the representation of PtX
  commodities for domestic use or export include: (1) Green ammonia production via
  the Haber-Bosch process, linking hydrogen, nitrogen (from air separation units),
  and electricity to produce NH3 (2) Green methanol synthesis from hydrogen and
  captured CO2 (3) Direct reduced iron (DRI) and hot-briquetted iron (HBI)
  production for green steel manufacturing, utilizing hydrogen-based direct reduction
  followed by electric arc furnace (EAF) steel production (4) Shipping sector defossilization
  through exogenous fuel choice among H2, NH3, methanol, and oil-based fuels (5)
  Usage of custom energy demand and industry production projections and three
  renewable generation profiles classes per region build on existing interfaces
  (6) A set of constraints to enforce hourly temporal matching of renewable
  electricity and battery discharge supply with electrolytic hydrogen and green
  hydrogen derivatives production, in line with emerging EU regulations.

  This set of constraints ensures compliance with emerging EU regulations on green
  hydrogen and renewable fuels of non-biological origin (RFNBOs).

  The mathematical representation of the hourly matching constraint, also
  considering battery flexibility, is as follows:

  \begin{align}
     & \sum_{g \in \mathcal{G}_{\text{RE}}}P_{g,t}+ \sum_{s \in \mathcal{S}_{\text{hydro}}}P_{s,t}^{\text{dispatch}}\notag                                                                 \\
     & \quad + \sum_{b \in \mathcal{B}_{\text{batt}}}\eta_{b}\cdot P_{b,t}\geq \sum_{\ell \in \mathcal{L}_{\text{H}_2}}P_{\ell,t}\quad \forall t \in \mathcal{T}\label{eq:hourly_matching}
  \end{align}
  \noindent
  where $\mathcal{G}_{\text{RE}}$ denotes the set of renewable generators (solar,
  wind, etc.), $\mathcal{S}_{\text{hydro}}$ the set of hydro storage units,
  $\mathcal{B}_{\text{batt}}$ battery dischargers, $\mathcal{L}_{\text{H}_2}$ electrolysis
  links, $P_{g,t}$ generation power, $P_{s,t}^{\text{dispatch}}$ hydro dispatch,
  $\eta_{b}$ battery efficiency, and $\mathcal{T}$ the set of hourly time steps.
  This hourly matching constraint is combined with an annual renewable matching
  constraint without batteries to prevent the use of fossil generators for
  charging the batteries. Furthermore, a green hydrogen allocation constraint is
  implemented to ensure that hydrogen allocated to export-oriented or downstream
  value chains is sourced from certified green hydrogen.

  \begin{align}
    \sum_{t \in \mathcal{T}}w_{t}\sum_{\ell \in \mathcal{L}_{\text{H}_2}}\eta_{\ell}\cdot P_{\ell,t}\geq \sum_{t \in \mathcal{T}}w_{t}\left( \sum_{k \in \mathcal{K}_{\text{conv}}}C_{k,t}^{\text{H}_2}+ \sum_{x \in \mathcal{X}_{\text{H}_2}}P_{x,t}\right) \label{eq:hydrogen_allocation_constraint}
  \end{align}

  \noindent
  where $\mathcal{K}_{\text{conv}}$ represents hydrogen conversion processes (e.g.
  Haber-Bosch, methanolisation), $\mathcal{X}_{\text{H}_2}$ direct hydrogen
  export links, $\eta_{\ell}$ electrolyzer efficiency, and
  $C_{k,t}^{\text{H}_2}$ the hydrogen consumption of conversion process $k$ at
  time $t$.

  \subsection{Potential area analysis}

  To identify locations suitable for renewable energy PtX deployment, the
  methodology developed in \cite{haeckner2024, zink2024} is used to construct
  techno-ecological feasibility maps for utility-scale solar PV, onshore wind,
  and PtX. A Boolean overlay of georeferenced exclusion layers integrating land-use,
  technological, and economic constraints mask unavailable sites and produce a base
  map of deployable potential. Land use exclusions include forest, urban areas,
  cropland, water bodies, permanent snow, and mixed land cover classes.
  Additional criteria were considered in order to delineate deployable areas. Firstly,
  projects must meet a minimum solar and wind implementation probability score (SIP/WIP)
  ratio of 2. Sites within 30 km of the electricity grid are considered to
  ensure integration into existing infrastructure and for PtX allocation, suitable
  areas are restricted to coastal regions with sea access or areas where at least
  80\% of the land is projected to experience low water stress
  \cite{WRI_Aqueduct} by 2050.

\begin{figure*}[t]
    \centering
    \begin{subfigure}
      {0.48\textwidth}
      \centering
      \includegraphics[width=\linewidth]{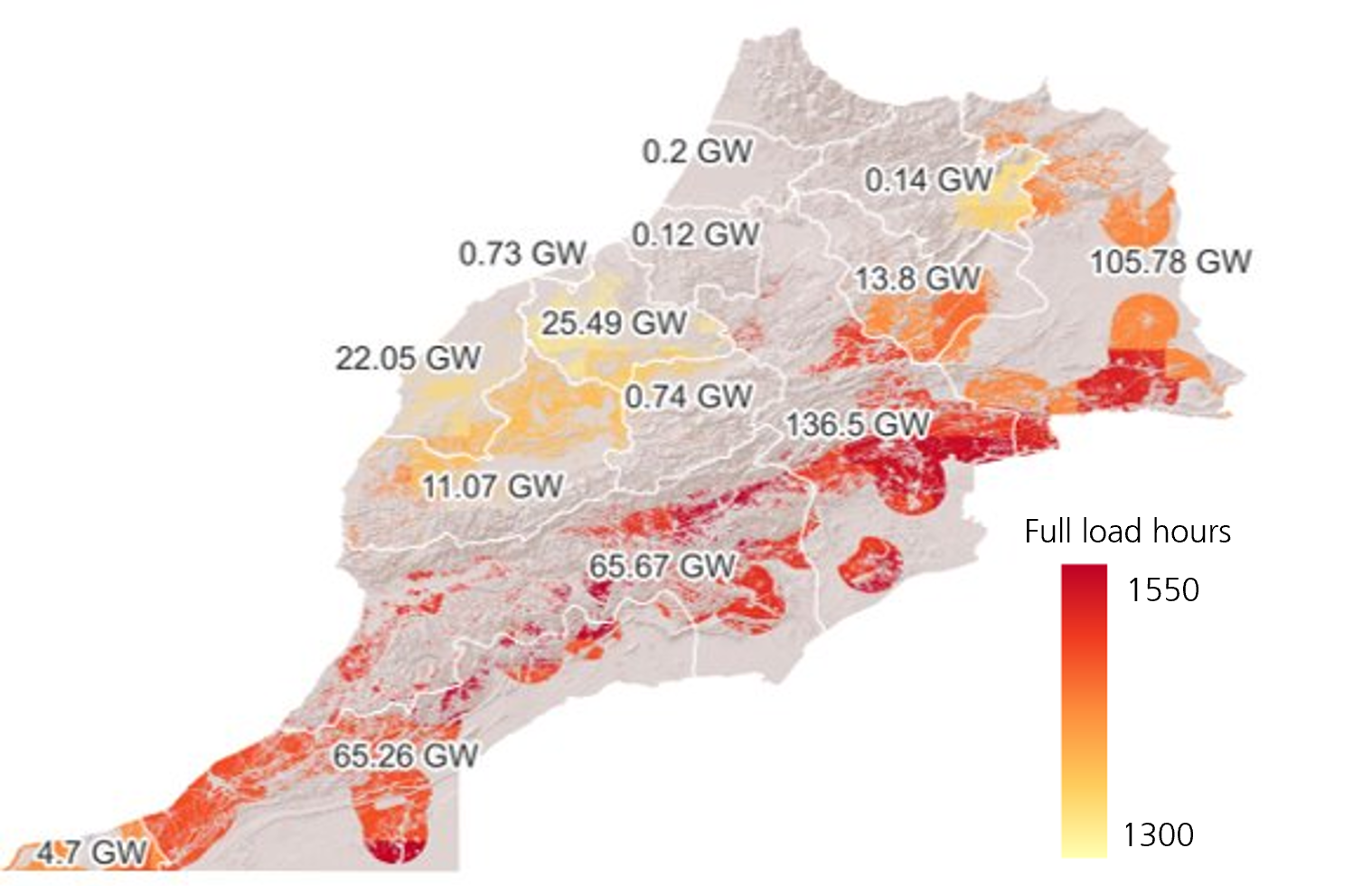}
      \caption{\textit{Regional distribution of solar PV full-load hours}}
      \label{fig:potential_area_pv}
    \end{subfigure}\hfill
    \vspace{0.5em} 
    \begin{subfigure}
      {0.48\textwidth}
      \centering
      \includegraphics[width=\linewidth]{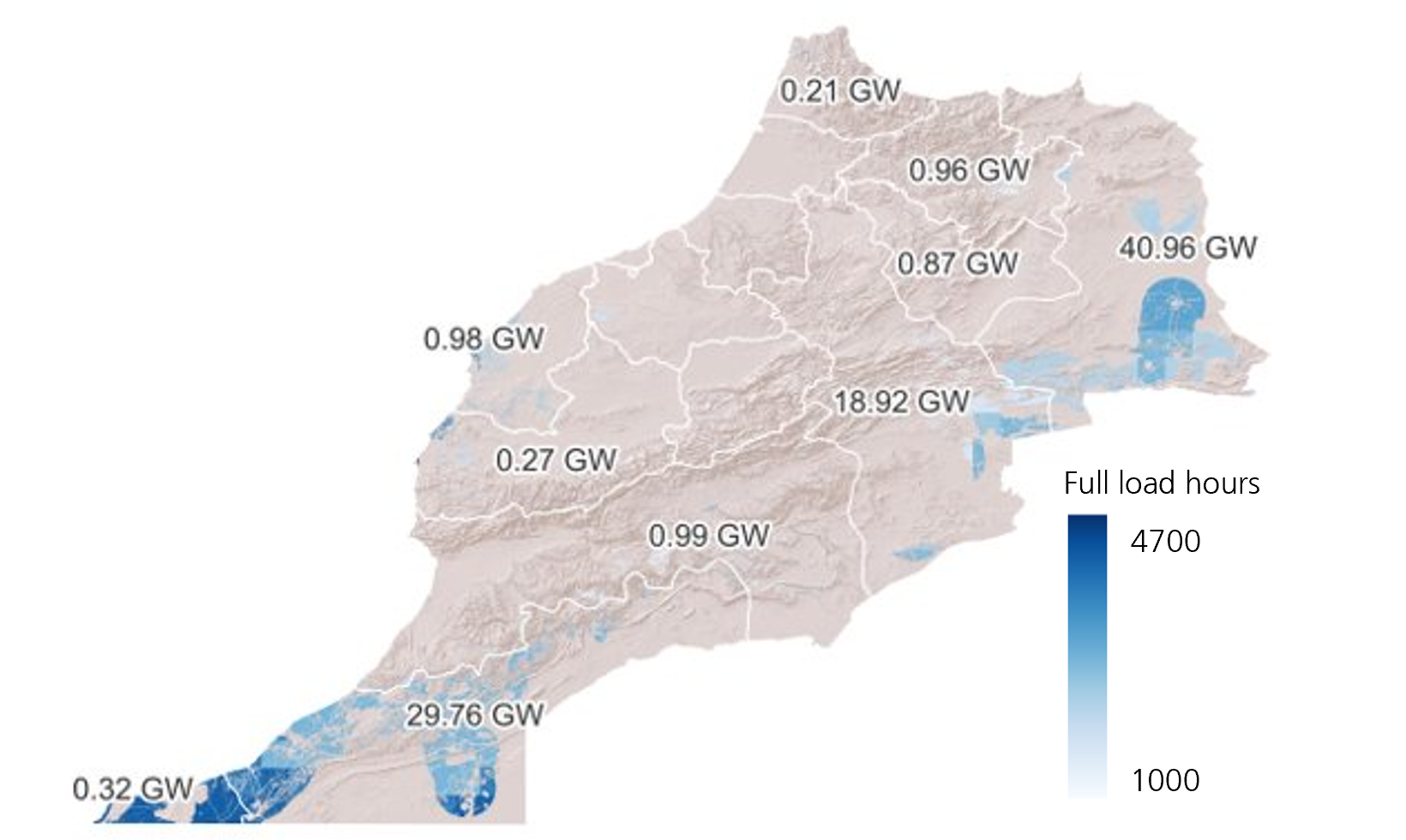}
      \caption{\textit{Regional distribution of onshore wind full-load hours}}
      \label{fig:potential_area_wind}
    \end{subfigure}\hfill
    \caption{\textit{Region-wise map of the potential areas and full load hours
    for onshore wind and solar-utility PV}}
    \label{fig:potential_area_analisis}
  \end{figure*}

  To illustrate the interim results of the potential area analysis, Figure
  \ref{fig:potential_area_analisis} presents the identified suitable locations. Following
  the application of the described methodology, a total available capacity of
  approximately 94.24 GW for onshore wind and 462.25 GW for solar PV was
  delineated. The analysis further indicates capacity factors reaching up to 53.7\%
  for onshore wind and 17.7\% for solar PV.

  \subsection{Renewable time series and sensitivity experiments}
  \label{sec:timeseries} For the multi-area energy system analysis, the spatial resolution
  consists of 14 regions (nodes) from the Global Administrative Areas (GADM) dataset,
  each representing a model node at which demand and supply constraints are
  enforced. At the regional level, we adopt the methodology in
  \cite{haeckner2024, zink2024}, using percentile-based clustering by capacity factor
  as the standard procedure, as it yields the clearest differentiation of site
  quality. This approach affords a more granular representation of high-quality wind
  and solar resource areas, often coincident with substantial available land, within
  each country. Specifically, the top 5\% of sites are designated Class 1 (95th-100th
  percentile), the next 10\% Class 2 (85th-95th percentile), and the remaining potential
  areas Class 3 (0th-85th percentile).

  For each node, representative generation profiles were constructed by
  weighting the contributions of the corresponding ERA5 grid cells according to
  their respective area. In combination with technology-specific power densities
  \cite{Pfennig.2023}, and under the assumption of wind and PV co-location
  within the same areas, this approach allows for the determination of maximum developable
  potentials for each cluster class. In order to refine the generation profiles,
  additional criteria were applied. Specifically, a minimum wind speed of 7 $m/s$
  at 150 m above sea level, corresponding to approximately 2300–2400 full load
  hours, was required for wind sites. Furthermore, the IEC Class 3 turbine configuration
  \cite{IEC61400} was employed for the time series simulations. Additionally, the
  weather year selection is based on nominal wind full load hours, given its
  higher variability nature compared to solar PV.

  Interpreting the relationships embedded in the model outcomes is challenging, given
  the number of interacting parameters. To identify the most influential drivers,
  we conducted a one-at-a-time sensitivity analysis. Among the different sources
  of uncertainty, weather variability, the weighted average cost of capital (WACC),
  and the pace of emissions reduction emerged as the most consequential.
  Sensitivity tests were therefore carried out on these parameters to evaluate
  their influence on the resulting technological allocation.

  \subsection{Sectoral demand pathways}
  \label{sec:demands} Constructing representative energy systems and plausible
  projections of future demand is a central challenge in the modelling workflow,
  given substantial socio-economic, environmental, and geopolitical uncertainties
  that shape scenario narratives \cite{Pypsasec2025}. For Morocco, we implement a
  hybrid demand-modelling approach that combines LEAP’s bottom-up apporach,
  which is implemented via the HYPAT Excel templates
  \cite{Muller2023_LEAP_Excel},\cite{LEAP_UserGuide_2005}; we integrated targeted
  literature review to construct national-level scenarios across three major end-use
  sectors: (i) \emph{residential} households; (ii) \emph{transport}—including
  road, rail, and domestic and international aviation and navigation; and (iii) \emph{industry}—covering
  iron and steel, chemicals and petrochemicals, non-metallic minerals, and paper
  and pulp production. Remaining sectors and subsectors-services, agriculture, and
  other manufacturing (mining and quarrying; food and tobacco; wood products; textiles
  and leather; construction; transport equipment; machinery)—are parameterised
  using the default demand trajectories from the PyPSA-Earth workflow
  \cite{Pypsasec2025}. Core data sets and baseline assumptions are compared with
  national scenarios and hydrogen strategies and benchmarked against global outlooks
  \cite{IEAInternationalEnergyAgency}, \cite{JRC_GECO_2024}. Scenario differentiation
  is achieved through sector-specific assumptions on activity levels, growth
  rates, technology diffusion, energy intensities, conversion efficiencies and fuel
  shares; detailed information on these parameters within the scenario pathways
  is provided in the following Section~\ref{sec:scenarios}. In general, final
  energy demand for sector $s$ in year $t$ follows the following standard end-use
  bottom-up formula:

  \begin{align}
    D_{s,c,t}\;=\; \sum_{u \in \mathcal{U}_s}A_{u,t}\,\cdot S_{u,t}\,\cdot I^{\mathrm{final}}_{u,t}\,\cdot \theta_{u,c,t}, \qquad \\
    E_{s,t}=\sum_{c \in \mathcal{C}}D_{s,c,t}, \label{eq:sector_by_carrier}
  \end{align}

  with total sector demand $E_{s,t}=\sum_{c \in \mathcal{C}}D_{s,c,t}$. Here, $u\in
  \mathcal{U}_{s}$ indexes end-uses within sector $s$; $c\in\mathcal{C}$ indexes
  energy carriers; $A_{u,t}$ is the activity driver (e.g., households, tonne-km);
  $S_{u,t}\in[0,1]$ is the saturation (share of activity using end-use $u$);
  $I^{\mathrm{final}}_{u,t}$ is the final-energy intensity per unit activity; and
  $\theta_{u,c,t}\in[0,1]$ is the carrier share mapping end-use $u$ to carrier $c$,
  with $\sum_{c\in\mathcal{C}}\theta_{u,c,t}=1$.

  To assess how emerging value chains reshape Morocco's energy system, all sectoral
  demands were held constant across scenarios, except for petrochemicals (ammonia
  and methanol), iron and steel, and the emerging green hydrogen market, which were
  allowed to vary due to their central role in the transition. Moderate
  electrification assumptions were applied in the transport sector to reflect
  gradual uptake of battery-electric vehicles across light- and heavy-duty road transport,
  buses, and two- and three-wheelers
  \cite{IEA_RoadTransport_2023, Khalili2019_GlobalTransport,
  Khalili2019_Supplement}, while projections for maritime and aviation sectors relied
  on \cite{IEA_AviationShipping_2023, DNV_MaritimeForecast2050_2024}. This
  generates an additional 6.1 TWh of electricity demand, equivalent to 8.7\% of
  domestic land transport energy use (70.1 TWh).

\begin{figure} 
    \centering
    \includegraphics[width=\columnwidth]{
      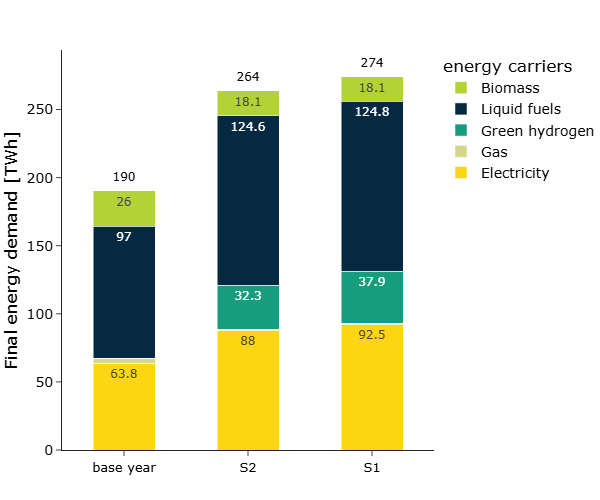
    } 
    \caption{\textit{Final energy demand by carrier in 2035 compared to the base
    year 2021. Note that the final energy demand for "Green hydrogen" is in TWh$H
    _{2}$.} }
    \label{fig:energy_demand_by_sector}
  \end{figure}

  The most substantial technological shift arises from increased adoption of air-conditioning
  systems, which drives electricity consumption from 15.7 TWh to 35.1 TWh, a 123\%
  increase, based on both country-specific data
  \cite{Krarti2019, JIHAD201885, ElHafdaoui2023, Jihad2016} and international
  sources
  \cite{Khalfallah_et_al_2016_AC_Maghreb, WorldBank_2014_CleanCooking_SSA,
  UN_DESA_PD_Household_Methodology_2022}
  (Figure~\ref{fig:energy_demand_by_sector}). Despite these electrification trends,
  fossil fuels remain dominant in the energy supply by 2035, driven primarily by
  demographic expansion and sustained economic growth from the Shared Socioeconomic
  Pathways (SSP)\cite{CSE_dataset_2023}, which collectively increase absolute energy
  demand.

  \subsection{Scenarios}
  \label{sec:scenarios}

  We construct two contrasting scenarios to reflect alternative strategic
  priorities and plausible market-formation pathways for Morocco through 2035, a
  mid-term horizon aligned with typical project cycles, network lead times, and announced
  investments expected to mature by that date, notably with a focus on hydrogen
  and its derivatives applications in the industry and only then on long-term
  for with applications in the heavy transport sector
  \cite{VespermannThielmann_HYPAT_2023,OCP2023, BakerMcKenzie_Morocco_Offer_2024}.

  The first scenario, \textit{Industry reallocation (S1)}, emphasizes domestic transformation
  and the strengthening of higher-value production processes in green value chains.
  It examines: (i) development of a green steel value chain oriented to both local
  and export demand; and (ii) ammonia aimed at full domestic autosupply for
  fertilizers, prioritizing energy security and resilience. Consistent with 
  process-integrated carbon streams from cement equipped with CCS (including
  biogenic co-processing where applicable) are treated as the binding carbon feedstock
  for methanol synthesis, such that methanol expansion is explicitly constrained
  by available CO/CO$_{2}$ from this source.

  The second scenario, \textit{Export opportunities (S2)}, prioritizes direct
  hydrogen exports to Europe supported by foreign investment. Infrastructure assumptions
  include potential hydrogen transport and export infrastructure expansion at existing
  ports. This framing allows us to contrast welfare, infrastructure, and market-integration
  effects between an inward-looking industrial strategy and an outward-looking
  export strategy\footnote{Export volumes are derived from a quadratic (degree-2)
  fit to near-term targets to capture accelerating growth.}.

  Table\ref{tab:scenario_volumes_summary} provides an overview of the annual
  production volumes for green hydrogen and its key derivatives under each
  scenario.

  \begin{table}[t]
    \centering
    \footnotesize
    \caption{Exogenous annual green hydrogen and derivatives production volumes
    per scenario.}
    \label{tab:scenario_volumes_summary}
    \begin{tabularx}
      {\columnwidth}{@{}l l r X@{}} \toprule \textbf{Commodities} & \textbf{Unit}
      & \textbf{Value} & \textbf{Reference} \\
      \midrule \multicolumn{4}{@{}l}{\textbf{Industry reallocation (S1)}} \\
      Steel & kt/y & 3,615.51 & \cite{Eurofer2024} \\
      HBI & kt/y & 1,039.01 & \cite{Eurofer2024} \\
      Ammonia (domestic)\textsuperscript{a} & TWh$_{\mathrm{NH_3}}$ & 13.26 & \\ 
      Methanol\textsuperscript{b} & TWh$_{\mathrm{MeOH}}$ & 6.46 & \cite{NH2_MA}
      \\
      [2pt] \multicolumn{4}{@{}l}{\textbf{Export opportunities (S2)}} \\
      Green hydrogen & TWh & 17.10 & \cite{NH2_MA} \\
      Steel & kt/y & 2,327.35 & \cite{Eurofer2024} \\
      Ammonia (domestic)\textsuperscript{a} & TWh$_{\mathrm{NH_3}}$ & 3.54 & \\ 
      Methanol\textsuperscript{b} & TWh$_{\mathrm{MeOH}}$ & 2.02 & \cite{NH2_MA}

      \\
      \bottomrule
    \end{tabularx}
    \vspace{0.25em}
    \parbox{\columnwidth}{\raggedright\footnotesize\textit{Notes:} \textsuperscript{a}
    LHV(NH$_{3}$)
    $= 5.17~\mathrm{MWh}_{\mathrm{NH_3}}/\mathrm{t}_{\mathrm{NH_3}}$. \; \textsuperscript{b}
    LHV(MeOH) $= 5.528~\mathrm{MWh}_{\mathrm{MeOH}}/\mathrm{t}_{\mathrm{MeOH}}$.}
  \end{table}

  \subsection{Welfare proxy calculations}
  To complement the optimisation outputs, we calculate a welfare proxy that
  captures the system-level value associated with domestic production in the new
  PtX value chains. The model provides marginal prices for hydrogen, ammonia, methanol
  and low-carbon steel and HBI, defined as the shadow value of supplying one
  additional unit of each product. These marginal prices reflect the system’s opportunity
  cost structure and therefore serve as consistent welfare weights within the optimisation
  framework.

  The welfare proxy is calculated by multiplying the marginal price of each product
  by its corresponding production volume:

  \begin{equation}
    W = \sum_{i \in \{H_2, NH_3, MeOH, HBI, steel\}}p_{i}^{\text{marg}}\cdot Q_{i}
    , \label{eq:welfare}
  \end{equation}

  where $p_{i}^{\text{marg}}$ denotes the model-derived marginal price of
  product $i$, and $Q_{i}$ its annual production volume in the optimised solution.
  The resulting indicator does not represent market revenue or profits; rather,
  it provides a system-consistent measure of the implicit welfare contribution
  associated with expanding domestic PtX-related industrial activity.




    \section{Results and discussion}
    


  This section discusses the outcomes of two contrasting strategies for Morocco’s
  energy transition, i.e., Industry Reallocation (S1) and direct hydrogen Export
  Orientation (S2), focusing on their implications for the hydrogen balance,
  cost-effectiveness of renewable energy deployment for the proposed scenarios
  and price volatility under uncertainty.

  The analysis moves beyond technology metrics to interpret the structural trade-offs
  that arise between domestic industrial transformation and export-oriented
  hydrogen development.

  Results are reported in terms of installed capacities and energy supply
  requirements for key technologies: onshore wind, utility-scale solar, PEM electrolysers,
  Haber-Bosch, HBI/DRI + EAF, and methanol synthesis, across the corresponding scenario
  combinations. In addition, complementary technologies, including desalination,
  direct air capture (DAC) and carbon capture and storage (CC and CCS), are
  included. We interpret the marginal costs of green steel, ammonia, methanol,
  and green hydrogen as proxies for wholesale market prices, noting that bilateral
  over-the-counter (OTC) agreements are beyond the scope of this modelling framework.

  \subsection{Final scenario demand pathways}
  \label{sec:results_supply_mix}

  By 2035, Morocco’s total final energy demand rises from $190$ TWh in 2021 to approximately
  $274.0$ - $263.7$ TWh, around 1.4 times Morocco’s current demand. Electricity demand
  reaches $92.5$ TWh (+45.0\% growth compare to base year) for S1 and $88. 0$ TWh
  (+37.9\%) for S2, driven primarily by green hydrogen production, with hydrogen
  electrolysis accounting for about 35\% of total generation, and broader sectoral
  electrification. In Figure \ref{fig:energy_demand_by_sector} it is possible to
  see similar overall magnitudes for hydrogen demand, however, the scenarios
  differ fundamentally in the allocation and use of it, which shapes both infrastructure
  requirements and economic outcomes.

  \subsection{Hydrogen market system balance}

  \begin{figure*}[t]
    \centering
    \begin{subfigure}
      {\columnwidth}
      \centering
      \includegraphics[width=\linewidth]{
        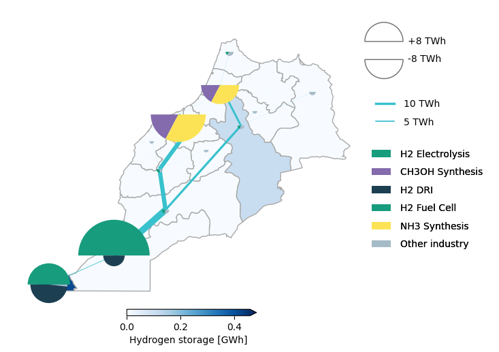
      }
      \caption{\textit{Green hydrogen hubs and storage requirements in S1}}
      \label{fig:hydrogen_market_s1}
    \end{subfigure}
    \vspace{0.5em}
    \begin{subfigure}
      {\columnwidth}
      \centering
      \includegraphics[width=\linewidth]{
        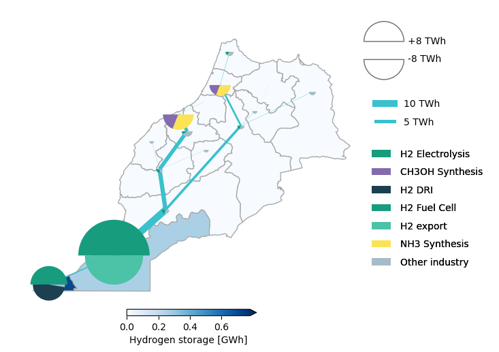
      }
      \caption{\textit{Green hydrogen hubs and storage requirements in S2}}
      \label{fig:hydrogen_market_s2}
    \end{subfigure}
    \caption{\textit{Regional distribution of green hydrogen market balance in
    Morocco for the analyzed scenarios. The upper circles indicate production,
    while the lower circles represent demand. All energy market flows are expressed
    in TWh$_{2}$.}}
    \label{fig:hydrogen_market}
  \end{figure*}

  The comparable hydrogen demand values for both scenarios ($32.3$ - $37.8$ TWh$_{\mathrm{H_2}}$)
  indicates that the volumes required to achieve near self-sufficiency in domestic
  ammonia production, covering $75\,\%$ of projected national demand by 2035, and
  for green steelmaking (roughly $26.0$ TWh$_{\mathrm{H_2}}$) are broadly
  aligned with the $25 .5$ TWh$_{\mathrm{H_2}}$ needed to meet Morocco’s export ambitions
  for green hydrogen and green methanol as outlined in the national hydrogen strategy.
  This proportionality reveals that the scale of Morocco’s domestic
  decarbonization potential is roughly equivalent to its targeted export
  capacity.

  The strategic challenge, therefore, is not only about increasing production
  but about the synergies with the allocation of renewable resources and the
  infrastructure required to transport green hydrogen or synthetic fuels to
  either to the off-taker locations, either export ports or industrial hubs.

  Figure \ref{fig:hydrogen_market_s1} illustrates the concentration of ammonia and
  HBI and steel production in Gharb–Chrarda–Béni Hssen and Greater Casablanca,
  consistent with existing industrial bases. In contrast, green steel production
  is allowed to be relocated to the South, to utilize the best renewable sites directly
  for producing green steel, which will be a highly competitive market. Experiments
  with national redistribution settings have shown that steel production in particular
  is sensitive to change. Future work can analyse this in depth. Utilizing the South
  is in line with the priorities set out in Morocco’s industrial strategy recently
  annouced in the moroccan offer. In Scenario 2 (Figure \ref{fig:hydrogen_market_s2}),
  Port Tan-Tan emerges as a key production hub, reflecting Morocco’s strategic
  positioning in international hydrogen markets. Other studies similarly identify
  Agadir and El-Jadida as viable alternatives with strong infrastructure and export
  potential. This underscores the need to plan for both retrofitting and
  capacity expansion, with an estimated 2 GW of green hydrogen production required
  to adequately supply the northern demand nodes.

  A total installed electrolyzer capacity of 3.8 $GW$ for S1 and 4.3 $GW$ (see
  Annex B for detailed figures) equivalent to nearly three times today’s global installed
  water-electrolyser capacity \cite{IEA2024_GlobalHydrogenReview}. Such a scale-up
  places substantial pressure on the expansion of electrolyser manufacturing
  capacity and highlights the need for accelerated industrial development across
  the value chain.


  \subsection{Strategic allocation of renewables}
\begin{figure*}[t]
    \centering
    \begin{subfigure}
      {0.48\textwidth}
      \centering
      \includegraphics[width=\linewidth]{
        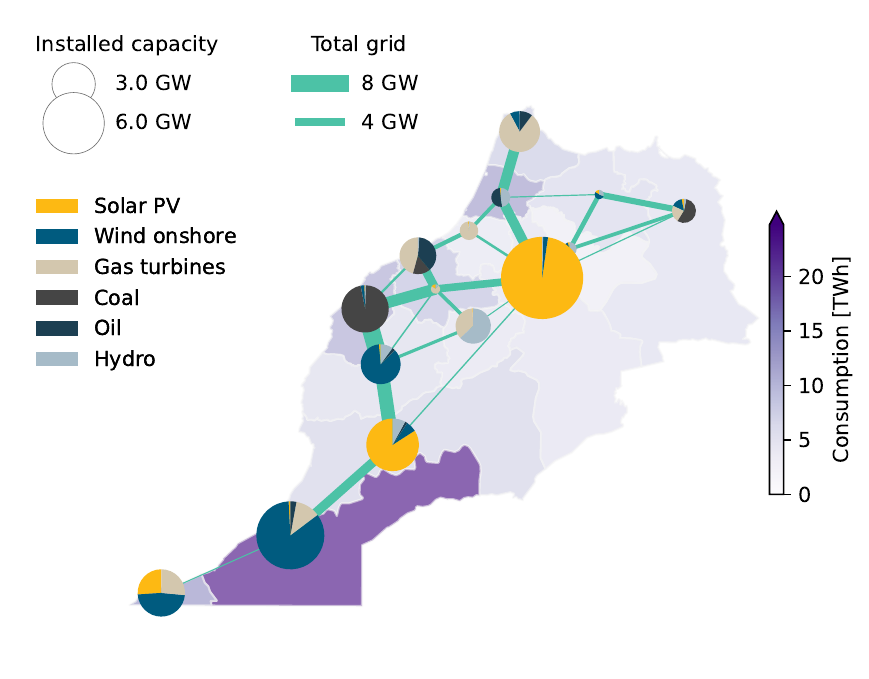
      }
      \caption{\textit{Regional distribution of electricity installed capacity
      by technology for S1}}
      \label{fig:sub1}
    \end{subfigure}\hfill
    \vspace{0.5em} 
    \begin{subfigure}
      {0.48\textwidth}
      \centering
      \includegraphics[width=\linewidth]{
        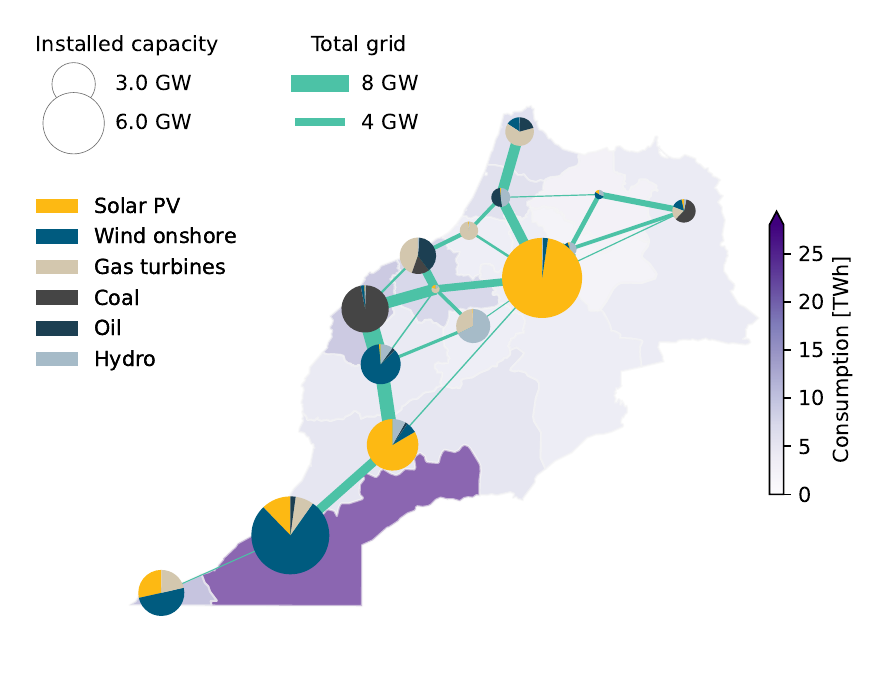
      }
      \caption{\textit{Regional distribution of electricity installed capacity
      by technology for S2}}
      \label{fig:sub3}
    \end{subfigure}\hfill
    \caption{\textit{Regional distribution of installed capacity by technology.The
    hydrogen consumption nodes and their intesity can be obtained from graphic, the
    region with the biggest demand is xx in both scenarios. The total grid
    capacity expansion required is about 8GW}}
    \label{fig:installed_capacities}
  \end{figure*}

  Consequently, balancing industrial self-sufficiency, foreign investment, and
  long-term resilience within a unified hydrogen framework requires a thorough
  analysis of the physical and technological feasibility of both strategies, as illustrated
  in Figure~\ref{fig:installed_capacities}.

  The similar spatial distribution of renewable energy and electrolysis capacities
  under both scenarios highlighting the presence of structurally robust investments
  areas that are resilient to shifts in PtX market orientation. Solar PV and
  onshore wind account for 63.9\% of ~52.8 GW and 67.5\% of ~49.8 GW of total
  renewable capacity in S1 and S2, respectively.

  The regional distribution further supports this robustness. In both scenarios,
  Meknès–Tafilalet consistently provides the highest solar PV potential with 9.7
  GW, over half of the national total (18 GW), while Guelmim–Es-Semara emerges as
  the strongest area for onshore wind with 9.9GW, roughly 63.5\% of total wind
  capacity (15 GW). These patterns remain stable between S1 and S2, suggesting
  that spatial resource quality, not market orientation, is the primary driver
  of optimal siting.

  A comparison with Morocco’s announced renewable project pipeline for 2030, based
  on the Global Solar Power Tracker \cite{GEM_GSPT_2024} and Global Wind Power Tracker
  \cite{GEM_GWPT_2024} (see Annex C), shows that the combined pipeline capacity
  already exceeds the optimized wind and solar capacities projected for 2035 in
  both modelled scenarios. This indicates that the scenarios fall well within the
  range of investments that are already underway, strengthening the credibility
  of the modeled expansion pathways. Moreover, the resulting capacity levels are
  consistent with Morocco’s NDC objective of reaching a 52\% share of renewables
  in the national generation mix.

  Finally, S1 requires an additional ~2 GW of pipeline capacity to transport hydrogen
  for green ammonia production and low-carbon steelmaking, reflecting the larger
  scale of domestic feedstock use in this pathway.

  Differences from other assessments may stem from two methodological features
  of this study: (i) the refined approach to identifying suitable deployment areas,
  which restricts installations to the highest-quality sites; and (ii) the use of
  percentile-based clustering of capacity factors to assign technologies into three
  distinct renewable resource classes, each associated with representative time-series
  profiles.

  \subsection{Sensitivity analysis on key parameters}

  To complement the core assessment, we evaluate the sensitivity of system outcomes
  to three key parameters: the stringency of $CO_{2}$ constraints, variations in
  financial conditions, captured through adjustments of the weighted average
  cost of capital (WACC) from the default 7.8\% to 6\%, climate-driven fluctuations
  in renewable generation profiles, represented by a low-wind ERA5 year (2016). Table
  \ref{tab:marginal_prices_summary} summarises the findings of the experiment.

  The dominance of WACC in shaping system outcomes underscores the extent to which
  Morocco’s PtX deployment pathways are governed by financial conditions rather than
  purely technical or climatic factors. Lower financing costs disproportionately
  favour capital-intensive technologies, such as PEM electrolysers, solar PV, and
  onshore wind, thereby reducing marginal prices across the system.

  In contrast, tightening the $CO_{2}$ constraint exerts a more moderate influence,
  This suggests that the system can reach near-zero-carbon outcomes without
  additional financial or regulatory incentives, reflecting the high competitiveness
  of renewables already embedded in the generation mix.

  The relatively minor effect of weather variability (+0.84\% in average) suggests
  that, within the assumed spatial distribution of renewables, the system is sufficiently
  diversified to absorb year-to-year fluctuations in resource availability.

  Overall, these results highlight that policy measures aimed at reducing
  investment risk and improving access to low-cost capital may be more effective
  in accelerating Morocco’s green hydrogen and PtX transition than adjustments
  to emissions targets alone.

  \begin{table*}
    [t]
    \centering
    \caption{Sensitivity of Marginal Prices to Key Uncertainties}
    \label{tab:marginal_prices_summary}
    \begin{tabular}{p{3cm} p{2cm} p{2.5cm} p{2.5cm} p{2.5cm} p{2.5cm}}
      \toprule Scenario                                                                                                              & Product  & Absolute value & Tight CO$_{2}$ & Low WACC & Low Wind \\
      \midrule \multirow{5}{*}{S1}                                                                                                   & H$_{2}$  & 111.3          & 103.7          & 91.7     & 100.3    \\
                                                                                                                                     & HBI      & 554.6          & 101.4          & 94.1     & 100.2    \\
                                                                                                                                     & Methanol & 164.5          & 100.4          & 93.6     & 102.8    \\
                                                                                                                                     & NH$_{3}$ & 154.5          & 102.6          & 91.7     & 100.8    \\
                                                                                                                                     & Steel    & 686.9          & 101.2          & 94.2     & 100.2    \\
      \midrule \multirow{5}{*}{S2}                                                                                                   & H$_{2}$  & 112.3          & 99.0           & 90.0     & 98.2     \\
                                                                                                                                     & HBI      & 556.2          & 99.1           & 93.4     & 99.6     \\
                                                                                                                                     & Methanol & 165.5          & 97.8           & 92.2     & 100.7    \\
                                                                                                                                     & NH$_{3}$ & 155.3          & 100.0          & 90.5     & 99.1     \\
                                                                                                                                     & Steel    & 688.5          & 99.2           & 93.7     & 99.7     \\
      \bottomrule \multicolumn{6}{l}{\footnotesize\textit{Notes: Units are in €2024/MWh for fuels and in €2024/t for steel and HBI.}}
    \end{tabular}
    \vspace{0.5em}
  \end{table*}

  \subsection{System costs and welfare proxy}
  Total annual system costs range from roughly €15.5 billion to €18.9 billion,
  with the highest-cost outcome observed in scenario S1. Although total system costs
  differ between scenarios roughly ±10–15\%, these expenditures should not be viewed
  solely as cost burdens. The scenario with the highest cost (S1) is also the
  one that delivers the deepest industrial diversification and the most
  substantial expansion in domestic value chains. Higher system costs therefore reflect
  strategic investment rather than inefficiency, as they enable local production
  of ammonia, methanol, HBI, and green steel, activities associated with
  productivity gains, reduced import exposure, and long-term competitiveness.

  Using the welfare-proxy derived from marginal prices, the additional domestic industrial
  activity generated in each scenario corresponds to an implicit welfare contribution
  of approximately €2.8 billion in S2 and €6.2 billion in S1 by 2035. These values
  are not market revenues but rather the shadow welfare associated with expanding
  local production of ammonia, methanol, HBI and green steel. Importantly, the scenario
  with the highest total system cost (S1) also yields the largest welfare contribution,
  reflecting the fact that more capital-intensive domestic value chains generate
  greater economic activity, reduce exposure to imported intermediates, and
  stimulate broader development spillovers. This indicates that interpreting pathways
  solely through a least-cost lens risks undervaluing the national welfare gains
  embedded in industrial diversification.


  Consequently, judging hydrogen strategies solely on least-cost criteria risks understating
  the broader socioeconomic benefits of emerging PtX industries. Planning
  frameworks should therefore integrate welfare-based indicators alongside cost
  metrics to better reflect national development priorities.

    \section{Conclusion and limitations}
    
  This paper assessed two contrasting strategies for Morocco's green hydrogen transition,
  Industry reallocation (S1) and Export opportunities (S2), integrating end-use electrification,
  PtX deployment, and regional infrastructure requirements into a unified national
  optimisation framework. The results highlight that both pathways rely on a
  similar order of magnitude of hydrogen demand by 2035 (32–38 TWh), indicating that
  Morocco's domestic decarbonisation needs are of the same structural magnitude as
  its export ambitions. The strategic question, therefore, shifts from how much
  hydrogen to produce to where value creation, industrial development, and long-term
  resilience should be prioritised.

  Comparison with Morocco’s announced renewable project pipeline for 2030 shows
  that planned solar and wind capacities already exceed the optimized capacity
  requirements for 2035 in both scenarios. This reinforces the feasibility of the
  proposed hydrogen pathways and provides technical validation for national hydrogen
  and renewable energy strategies, which often lack prior modelling evidence.

  Across scenarios, the model identifies structurally robust regions for large-scale
  renewable deployment, notably Meknès-Tafilalet for solar PV and Guelmim-Es-Semara
  for onshore wind. These results are resilient to uncertainties in weather variability
  and $CO_{2}$ constraints, but highly sensitive to financial conditions. The dominance
  of the WACC in shaping system outcomes underscores that Morocco's hydrogen trajectory
  is governed less by technical potential and more by access to low-cost capital.
  Policy measures reducing investment risk may therefore accelerate deployment
  more effectively than additional regulatory tightening.

  A central contribution of this study is its move beyond a purely cost-minimising
  interpretation of pathways. While total system costs lie in the range of €15.5–18.9
  billion, the highest-cost scenario (S1) is also the one that generates the largest
  welfare contribution. Using marginal prices as a welfare proxy, S1 yields approximately
  €6.2 billion of implicit welfare associated with domestic ammonia, methanol,
  HBI, and green steel production, compared with €2.8 billion in S2. These
  figures do not represent revenues but capture the broader economic value of
  establishing domestic value chains reduced import dependence, local industrial
  activity, and development spillovers. This finding highlights a crucial policy
  insight: least-cost pathways are not necessarily the ones that maximise
  national welfare.

  The results therefore point to a structural trade-off: export-led strategies can
  mobilise international investment and align with emerging hydrogen markets,
  but domestic industrial reallocation can generate higher welfare and more
  resilient socio-economic outcomes. Planning for Morocco's hydrogen future must
  balance these dimensions, acknowledging that a narrow optimisation perspective
  may undervalue long-term industrial capability formation.

  Taken together, the results suggest that Morocco’s hydrogen strategies are internally
  coherent and technically feasible under cost-optimal pathways. However, hydrogen
  deployment is not merely an energy transition issue; it is a broader
  development choice. The central challenge is less about producing green hydrogen
  at scale than about harnessing it to support industrial diversification,
  economic resilience and social welfare gains by 2035.

  It is beyond the scope of the study to perform a in-depth impact assessment of
  different socio-economic or technical constraints for moving industry capacities
  nationally. Reallocating or building new industry cluster in remote but
  renewable rich areas imposes multiple challenges which need to be further
  inspected in future work. On the model side, key limitations include: (1) Ignoring
  costs for port infrastructure and the interregional transport of ammonia,
  methanol or steel; (2) No representation of the production of ammonia downstream
  products, such as urea or other specific fertilizers; (3) Data gaps in technology
  or energy carrier splits and regional distribution factors for subsectoral and
  branch-specific end-uses; (4) Exclusion of rebound and price-elasticities; (5)
  Potential of cooperation with neighbouring countries in the Maghreb region;
  and (6) Abstraction from sub-hourly flexibility, detailed siting constraints and
  stochastic uncertainties. Future work can incorporate those aspects, such as endogenous
  port terminal capacity expansion for ammonia and other export value chains,
  and utilize stochastic optimization to account for uncertain future market
  potentials of different PtX products.

    \section{Acknowledgements and disclaimer}
    The authors gratefully acknowledge funding from the H2Global meets Africa
    project (FKZ:03SF0703B), supported by the German Federal Ministry of Research,
    Technology and Space (BMFTR). We also extend our gratitude to the PyPSA meets
    Earth community and PyPSA-Earth contributors for their valuable support,
    contributions, and the collaborative environment that enriched this work.

    This study uses administrative boundaries from GADM.org. The boundaries used
    do not represent the authors' opinion on the legal status of any country,
    territory, city or area, or its authorities.

    \section*{References}
    \bibliographystyle{IEEEtranN}
    \bibliography{reference}

    \section*{Appendix}
    \appendix
    
    \subsection*{Appendix A: Morocco's model nodes}
    The administrative regions considered as nodes in the model are based on GADM
    Level 1 (2015) as described in Table \ref{tab:region_mapping}.
    \begin{table}[H]
        \caption{Morocco's regions code and name mapping}
        \label{tab:region_mapping}
        \begin{tabular}{ll}
            \toprule GADM & Region Name                      \\
            \midrule MA 1 & Chaouia - Ouardigha              \\
            MA 2          & Doukkala - Abda                  \\
            MA 3          & Fès - Boulemane                  \\
            MA 4          & Gharb - Chrarda - Béni Hssen     \\
            MA 5          & Grand Casablanca                 \\
            MA 6          & Guelmim - Es-Semara              \\
            MA 7          & Laâyoune - Boujdour - Sakia El H \\
            MA 8          & Marrakech - Tensift - Al Haouz   \\
            MA 9          & Meknès - Tafilalet               \\
            MA 10         & Oriental                         \\
            MA 11         & Rabat - Salé - Zemmour - Zaer    \\
            MA 12         & Souss - Massa - Draâ             \\
            MA 13         & Tadla - Azilal                   \\
            MA 14         & Tanger - Tétouan                 \\
            MA 15         & Taza - Al Hoceima - Taounate     \\
            \bottomrule
        \end{tabular}
    \end{table}

    \subsection*{Appendix B: Additional optimazation results}
    \begin{figure}[t]
        \centering
        \begin{subfigure}
            {\columnwidth}
            \centering
            \includegraphics[width=\linewidth]{
                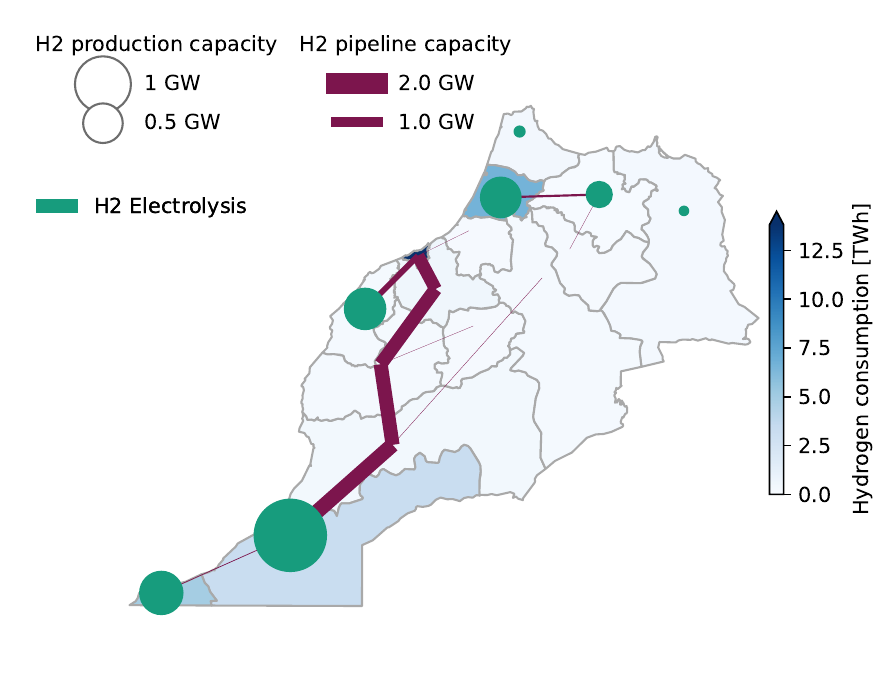
            }
            \caption{\textit{Green hydrogen hubs and storage requirements in S1.}}
            \label{fig:hydrogen_market_s1}
        \end{subfigure}
        \vspace{0.5em}
        \begin{subfigure}
            {\columnwidth}
            \centering
            \includegraphics[width=\linewidth]{
                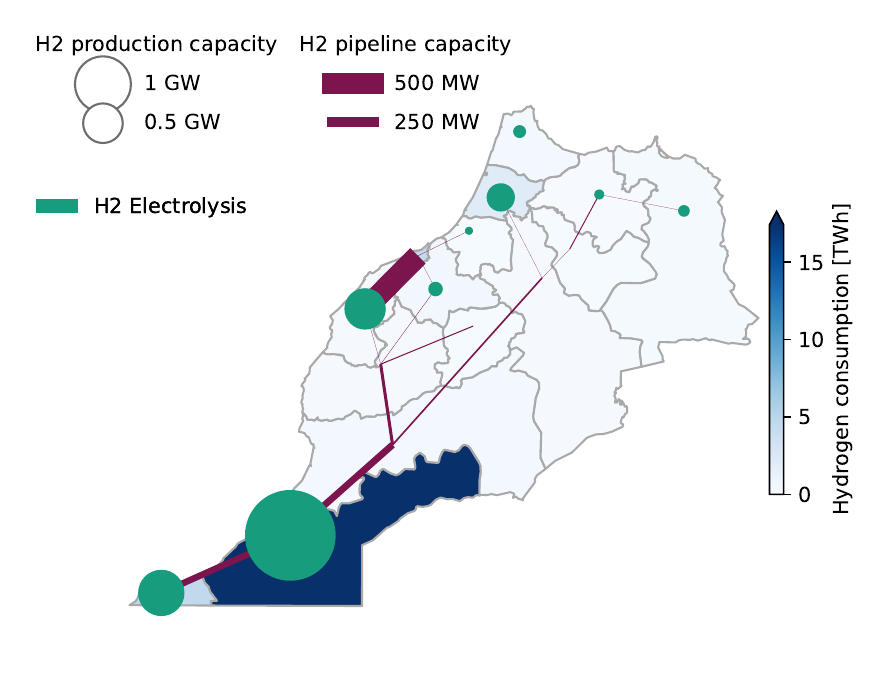
            }
            \caption{\textit{Green hydrogen hubs and storage requirements in S2.}}
            \label{fig:hydrogen_market_s2}
        \end{subfigure}
        \caption{\textit{Regional distribution of green hydrogen market balance
        in Morocco for the analyzed scenarios. All energy market flows are expressed
        in TWh$_{2}$.}}
        \label{fig:hydrogen_market}
    \end{figure}
    \newpage
    \subsection*{Appendix C: }
    \begin{figure}[t]
        \centering
        \begin{subfigure}
            {\columnwidth}
            \centering
            \includegraphics[width=\linewidth]{
                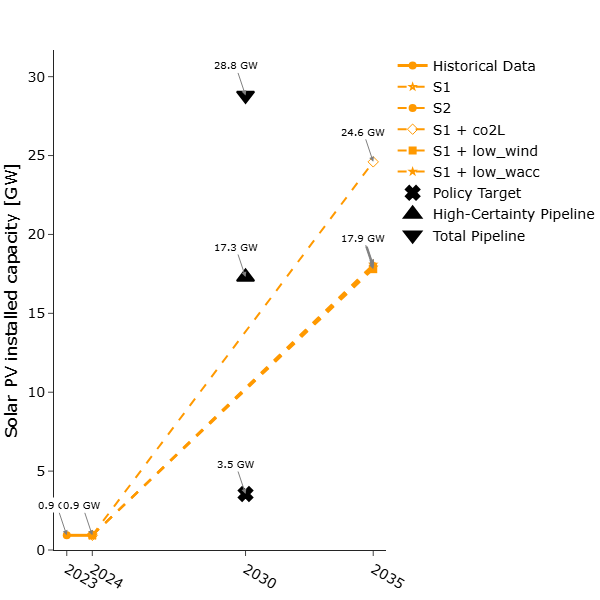
            }
            \caption{\textit{Total historical installed capacity for solar PV in
            Morocco, combined with the results of optimization and future announced
            projects.}}
            \label{fig:hydrogen_market_s1}
        \end{subfigure}
        \vspace{0.5em}
        \begin{subfigure}
            {\columnwidth}
            \centering
            \includegraphics[width=\linewidth]{
                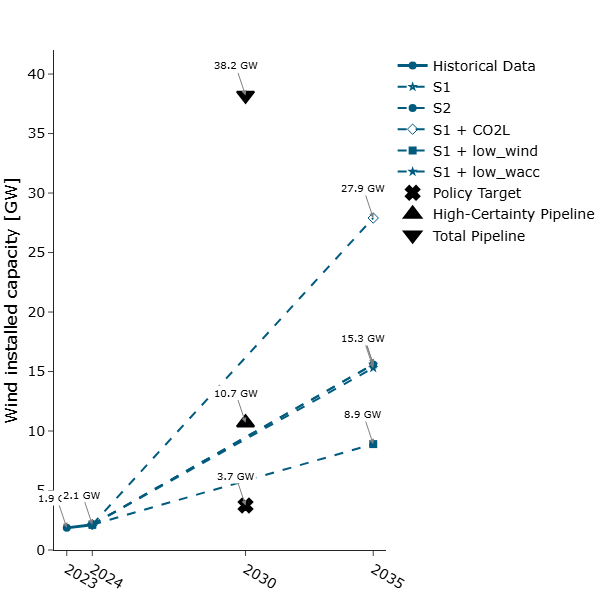
            }
            \caption{\textit{Total historical installed capacity for onshore
            wind in Morocco, combined with the results of optimization and future
            announced projects}}
            \label{fig:hydrogen_market_s2}
        \end{subfigure}

        \label{fig:hydrogen_market}
    \end{figure}

\end{document}